\documentclass[iop]{emulateapj}
\shorttitle{A Massive UCD in M60}
\shortauthors{Strader \etal~}
\def\etal{{\it et al.}}
\def\kms{\,km~s$^{-1}$}

\usepackage{subfigure}
\usepackage{amsmath}
\usepackage{graphics}

\begin{document}

\title{The Densest Galaxy}

\author{Jay Strader\altaffilmark{1}, Anil C.~Seth\altaffilmark{2}, Duncan A.~Forbes\altaffilmark{3}, Giuseppina Fabbiano\altaffilmark{4}, Aaron J.~Romanowsky\altaffilmark{5,6}, Jean P.~Brodie\altaffilmark{6}, Charlie Conroy\altaffilmark{7}, Nelson Caldwell\altaffilmark{4}, Vincenzo Pota\altaffilmark{3}, Christopher Usher\altaffilmark{3}, Jacob A.~Arnold\altaffilmark{6}}

\email{strader@pa.msu.edu}

\altaffiltext{1}{Department of Physics and Astronomy, Michigan State University, East Lansing, Michigan 48824}
\altaffiltext{2}{University of Utah, Salt Lake City, UT 84112}
\altaffiltext{3}{Centre for Astrophysics \& Supercomputing, Swinburne University, Hawthorn, VIC 3122, Australia}
\altaffiltext{4}{Harvard-Smithsonian Center for Astrophysics, Cambridge, MA, 02138}
\altaffiltext{5}{Department of Physics and Astronomy, San Jos\'e State University, San Jose, CA 95192}
\altaffiltext{6}{University of California Observatories/Lick Observatory, Santa Cruz, CA 95064}
\altaffiltext{7}{Department of Astronomy \& Astrophysics, University of California, Santa Cruz, CA 95064}

\begin{abstract}

We report the discovery of a remarkable ultra-compact dwarf galaxy around the massive Virgo elliptical galaxy NGC 4649 (M60), which we term M60-UCD1. With a dynamical mass of $2.0\times10^{8} M_{\odot}$ but a half-light radius of only $\sim24$ pc, M60-UCD1 is more massive than any ultra-compact dwarfs of comparable size, and is arguably the densest galaxy known in the local universe. It has a two-component structure well-fit by a sum of S\'ersic functions, with an elliptical, compact ($r_h=14$ pc; $n\sim3.3$) inner component and a round, exponential, extended ($r_h=49$ pc) outer component. \emph{Chandra} data reveal a variable central X-ray source with $L_X\sim10^{38}$ erg s$^{-1}$ that could be an active galactic nucleus associated with a massive black hole or a low-mass X-ray binary. Analysis of optical spectroscopy shows the object to be old ($\ga10$ Gyr) and of solar metallicity, with elevated [Mg/Fe] and strongly enhanced [N/Fe] that indicates light-element self-enrichment; such self-enrichment may be generically present in dense stellar systems. The velocity dispersion ($\sigma\sim70$ \kms) and resulting dynamical mass-to-light ratio ($M/L_V=4.9\pm0.7$) are consistent with---but slightly higher than---expectations for an old, metal-rich stellar population with a Kroupa initial mass function. The presence of a massive black hole or a mild increase in low-mass stars or stellar remnants is therefore also consistent with this $M/L_V$. The stellar density of the galaxy is so high that no dynamical signature of dark matter is expected. However, the properties of M60-UCD1 suggest an origin in the tidal stripping of a nucleated galaxy with $M_B~\sim-18$  to $-19$. 

\end{abstract}


\section{Introduction}

Objects with sizes and masses between those of globular clusters and compact ellipticals ($r_h\sim10$--100 pc; $M_V\sim-9$ to $-14$) were first discovered in spectroscopic surveys of galaxy clusters (Hilker \etal~1999; Drinkwater \etal~2000). They were quickly dubbed ``ultra-compact dwarf" galaxies (UCDs), even though their galaxian nature was unclear. Large populations of UCDs have been discovered in Fornax, Virgo, and other galaxy clusters, as well as in group and field environments---see reviews in Chilingarian \etal~(2011), Norris \&  Kannappan (2011), and Brodie \etal~(2011).

UCD formation scenarios have coalesced around two poles: star cluster or galaxy. In the former scenario, UCDs form the massive end of the normal sequence of globular clusters (Mieske \etal~2012). Further, if some star clusters form in gravitationally-bound complexes, these can merge to make objects that are larger and more massive than single clusters (Br{\"u}ns \etal~2011).

Alternatively, UCDs could be galaxies that formed in individual dark matter halos---either ``in situ", as unusual, extremely compact galaxies---or as the products of tidal stripping of more massive progenitor galaxies (e.g., Drinkwater \etal~2003).

A reasonable synthesis of these scenarios may be that the least-massive ``UCDs", with $\sim10^6 M_{\odot}$, are largely star clusters, while the most massive objects ($\ga10^8 M_{\odot}$) are galaxies, or the tidally stripped remnants thereof. At intermediate masses both star clusters and galaxies may co-exist (e.g., Norris \& Kannappan 2011; Brodie \etal~2011).

There is more at stake than the natural desire to understand these novel stellar systems. If a significant fraction of UCDs contain dark matter, then they form a populous class that must be included in counts of subhalos for comparisons to cosmological theory. Further, if some UCDs are formed by tidal stripping, their chemical and structural properties help trace galaxy transformation.

Here we report the discovery of an extraordinary UCD around the Virgo elliptical NGC 4649 (M60). It has a half-light radius of 24 pc but a stellar mass of $2 \times 10^{8} M_{\odot}$, giving it the highest surface density of any galaxy in the local universe. We also present evidence that this UCD may contain a central supermassive black hole.

\section{Data}

\subsection{Imaging}

We discovered M60-UCD1  in the \emph{Hubble Space Telescope}/Advanced Camera for Surveys imaging of Strader \etal~(2012). We have a single orbit of imaging split between $F475W$ and $F850LP$ (hereafter $g$ and $z$). M60-UCD1 is located at (R.A., Dec.) = (190.8999, 11.5347) in decimal J2000 coordinates. This is at a projected distance of only $\sim6.6$ kpc from the center of M60 (Figure 1; assuming a distance of 16.5 Mpc; Blakeslee \etal~2009). No mention is made of M60-UCD1 in previous Virgo surveys, including the ACS Virgo Cluster Survey (C{\^o}t{\'e} \etal~2004). It is present in the SDSS DR7 photometric catalog (Abazajian \etal~2009) as J124335.96+113204.6, and was classified by Simard \etal~(2011) as a background galaxy.

\begin{figure*}[ht!]
\centering
\epsscale{1.15}
\plotone{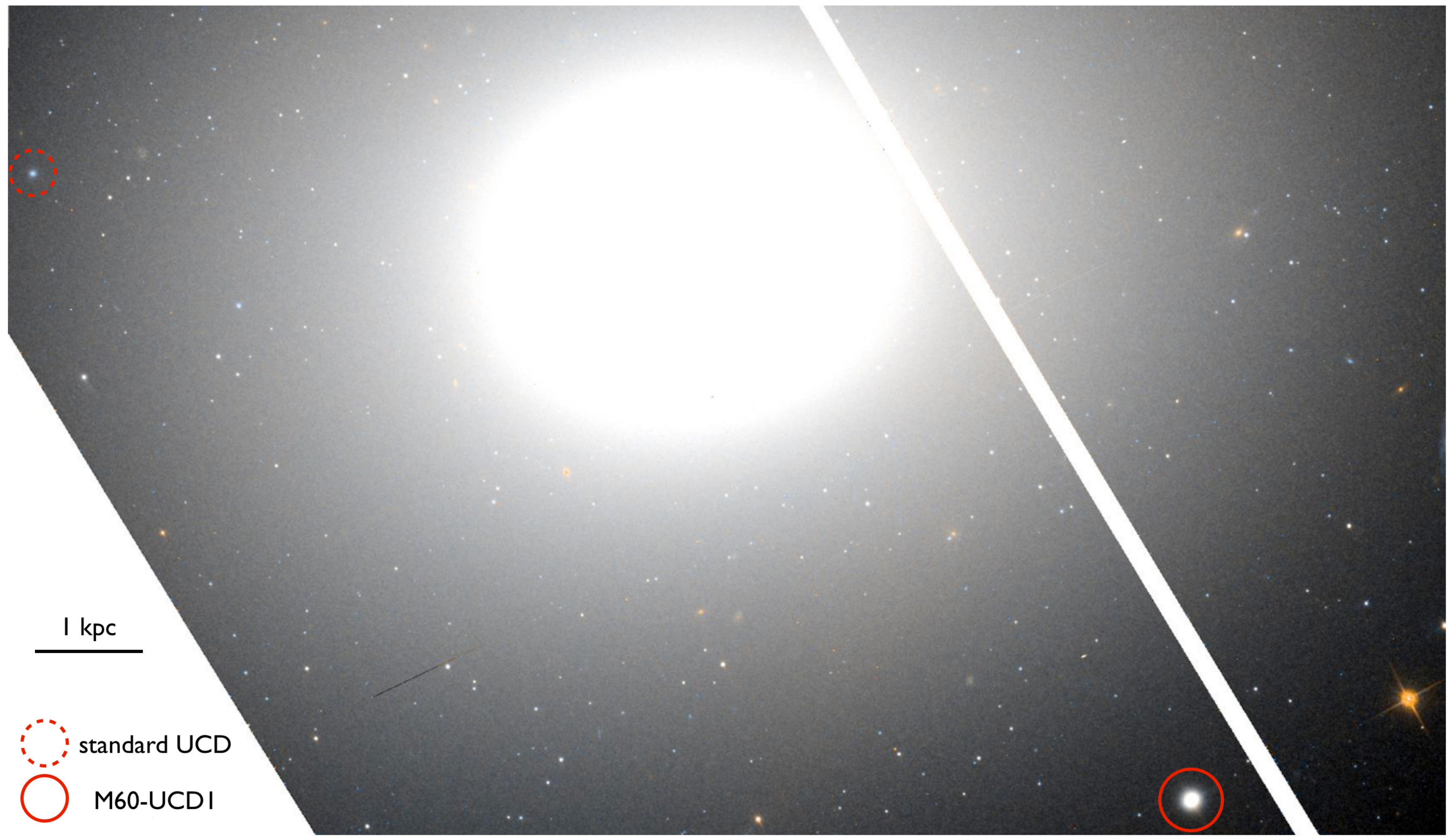}
\caption{$HST$/ACS color image of the central region of M60, showing the location of M60-UCD1 (solid circle). A typical UCD (A32, $\sim3\times10^6 M_{\odot}$; Strader \etal~2012) is also marked (dashed circle) for reference.}
\label{pic}
\end{figure*}

\subsection{Spectroscopy}

A spectrum of M60-UCD1 was obtained on the night of 17 January 2012 with Keck/DEIMOS (Faber \etal~2003), utilizing the 1200 l/mm grating centered at 7800 \AA\ and a 
1\arcsec\ slit (resolution $\sim1.5$ \AA). We obtained three 30-min exposures in 0.8\arcsec\ seeing. Using the {\tt spec2d} pipeline (Cooper \etal~2012), the spectra were extracted, calibrated, and combined in the standard manner to produce a final one-dimensional spectrum. 

To improve stellar population constraints, further spectroscopy was undertaken with MMT/Hectospec (Fabricant \etal~2005) on 16 May 2012, using the 270 l/mm grating with wavelength coverage from 3700 to 9100 \AA\ and 5 \AA\ resolution. Three 20-min exposures were taken in 0.9\arcsec\ seeing. These Hectospec data were pipeline-reduced in a standard manner as described in Mink \etal~(2007).

\section{Analysis and Results}

\subsection{Imaging}

Aperture photometry of M60-UCD1 gives a total integrated magnitude of $z=15.86\pm0.02$ and $g=17.40\pm0.02$, yielding  $g-z=1.54\pm0.03$ (the $g$ and $z$ magnitudes in this paper are AB). The measured half-light radius (see below) is $r_e=24.2\pm0.5$ pc. The inferred total luminosities are: $L_g=(3.26\pm0.06)\times10^7L_{\odot}$; $L_z=(7.88\pm0.14)\times10^7L_{\odot}$; and $L_V=(4.12\pm0.08)\times10^7L_{\odot}$. With $M_V=-14.2$, M60-UCD1 is the most luminous UCD known (see \S 4.1 for further discussion). 

We fit the optical imaging with one and two-component elliptical S\'ersic models. These fits, shown in Figure 2, were performed by fitting two-dimensional models convolved with an empirical point spread function (PSF) using a custom software package as described in Seth \etal~(2006). The PSF ($10\times$ subsampled) was derived from point sources in the images. The fits were performed on a 5\arcsec$\times$5\arcsec\ image centered on the UCD, with the background galaxy gradient from M60 itself removed. The fits are not very sensitive to the fitting box size.

\begin{figure}
\centering
\epsscale{1.0}
\plotone{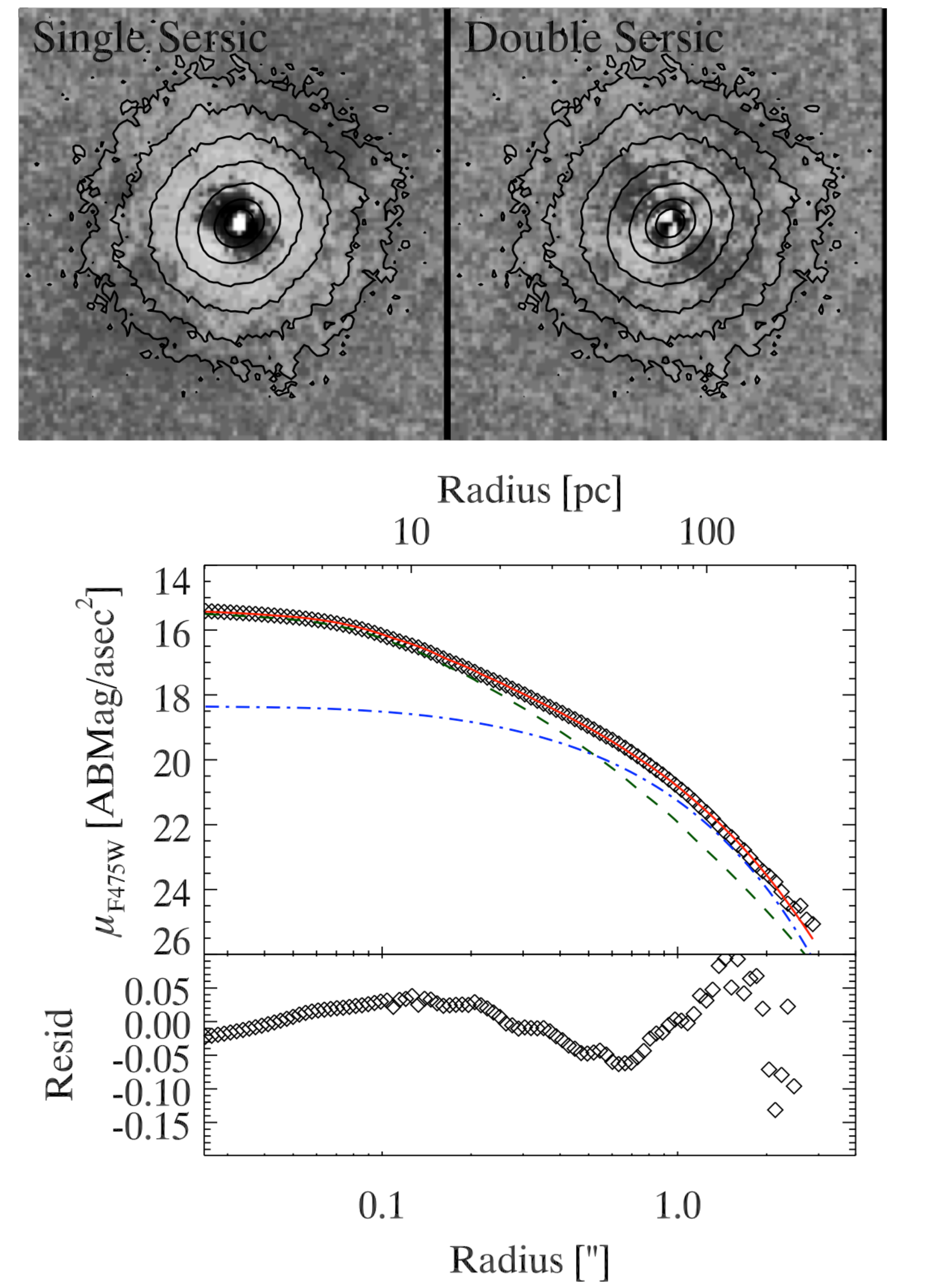}
\caption{{\bf Top panel}: The two-dimensional residuals of the best-fitting single S\'ersic (left) and double S\'ersic (right) fits. Contours show the $g$ surface brightness at $\mu_{\rm g}=17$ to 23 mag/arcsec$^2$. {\bf Bottom panel}: One-dimensional profile showing the results of our two-dimensional fits to the $g$ surface brightness profile of M60-UCD1. The fit is shown in the top panel and the residual (data minus model, units of mag/arcsec$^2$) in the bottom panel. Open black diamonds show the data and residuals, green dashed and blue dot-dashed lines the inner and outer components, and solid red line the sum. All fits were performed in two dimensions; these profiles are for display only.}
\label{sbfig}
\end{figure}

From the residual map (Figure 2), it is clear that a single S\'ersic component provides a poor fit.  In particular, the ellipticity of M60-UCD1 becomes more circular at larger radii, leaving a residual along the minor axis. The radial shape of the surface brightness profile is also poorly fit.

However, using a two-component S\'ersic model, a very good fit ($\chi^{2}_{\nu}$=1.07 in $g$) is obtained.  Table~1 gives the parameters for the single and double S\'ersic $g$ band fits and the double S\'ersic $z$ fits.  For the two-component fits the S\'ersic parameters $n$ and $r_e$ are very similar between the filters.  Because the $g$ band provides a much better fit (probably due to PSF modeling issues in $z$), all structural values cited are from the $g$ fits.

For the best-fit two-component model, the inner component is compact ($r_e=14$ pc), with modest ellipticity ($\epsilon=0.25$), and has about 58\% of the total luminosity of M60-UCD1. The outer component is more extended ($r_e=49$ pc), round, and with a nearly exponential profile ($n\sim1.2$). The overall half-light radius is $r_e=24.2\pm0.5$ pc, empirically measured using the deconvolved $g$ surface brightness profile. This value is similar to the radius derived from the single component S\'ersic fit; within this radius we estimate $L_V\sim2.1\times10^7 L_{\odot}$.  

\begin{deluxetable*}{lcccccccc}
\centering
\tablecaption{Surface Brightness Profile Fits \label{tab:datz}}
\tablehead{ Component &  $\chi^{2}/\nu$  & Total\tablenotemark{a}                     & Luminosity                                & $\mu_{e}$                      & $r_e$                 & $n$                       & $b/a$                           & P.A. \\
                                             &                            & (AB mag)                   &   ($L_\odot$)                            & (mag/arcsec$^{2}$)       & (pc)                    &                             &                                     & (deg)  }

\startdata

$g$ Single               & 2.33                               & 17.51 &   $2.92\times10^7$    &  $18.32\pm0.01$         &  $27.3\pm0.1$ & $3.53\pm0.01$ & $0.870\pm0.001$  & $-49.2\pm0.3$ \\ 
$g$ Double (inner)  & 1.07\tablenotemark{b}  & 18.11 &  $1.69\times10^7$     &  $17.35\pm0.08$         &  $14.3\pm0.7$ & $3.32\pm0.08$ & $0.750\pm0.004$  & $-47.0\pm0.3$ \\   
$g$ Double  (outer)  & \nodata                          & 18.46 &   $1.22\times10^7$    &   $20.13\pm0.06$         & $49.1\pm0.5$ &  $1.18\pm0.03$ &  $0.964\pm0.005$ &  $-10\pm9$ \\
$z$ Double (inner)  & 2.03\tablenotemark{b}  &   16.55 &   $4.17\times10^7$    &   $15.77\pm0.06$         & $14.6\pm0.4$ &  $3.28\pm0.06$ &  $0.708\pm0.004$ &  $-49.4\pm0.3$ \\
$z$ Double (outer)  & \nodata                          & 16.89 &    $3.05\times10^7$    &   $18.57\pm0.04$         & $50.4\pm0.5$ &  $1.14\pm0.02$ &  $0.930\pm0.007$ &  $109\pm3$ \\

\enddata
\tablenotetext{a}{Errors are dominated by the sky determination and are $<0.05$ mag.}
\tablenotetext{b}{$\chi^{2}/\nu$ applies to both components.}
%tablecomments{The double S\'ersic $g$ profile is the best fitting profile.  (a) Errors on total magnitudes are dominated by the sky determination and are $< 0.05$ mag.}
\end{deluxetable*}

%\begin{deluxetable}{|l|c|c|c|}
%\tablecaption{Surface Brightness Profile Fits \label{tab:dat1}}
%\hline
%Parameter & $g$ Single & {\bf $g$ Double} & $z$ Double\\
 %& S\'ersic & {\bf S\'ersic} & S\'ersic \\
%\hline
%Red.~$\chi^2$ & 2.33 & 1.07 & 2.03 \\
%\hline
%& Single S\'ersic & \multicolumn{2}{|c|}{Inner S\'ersic}\\
%\hline
%AB Mag.\tablenotemark{a}& 17.51 & 18.11 & 16.55 \\
%$L/L_\odot\tablenotemark{a}$ & 2.92$\times$10$^7$ & 1.69$\times$10$^7$ & 4.17$\times$10$^7$ \\
%$\mu_{e}$ & 18.32$\pm$0.01 & 17.35$\pm$0.08 & 15.77$\pm$0.06\\
%$r_e$ [pc] & 27.3$\pm$0.1 & 14.3$\pm$0.7 & 14.6$\pm$0.4\\
%$n$ & 3.53$\pm$0.01 & 3.32$\pm$0.08 & 3.28$\pm$0.06 \\
%$b/a$ & 0.870$\pm$0.001 & 0.750$\pm$0.004 & 0.708$\pm$0.004 \\
%PA [$^\circ$] & -49.2$\pm$0.3 & -47.0$\pm$0.3 & -49.4$\pm$0.3 \\
%\hline
%& & \multicolumn{2}{|c|}{Outer S\'ersic}\\
%\hline
%AB Mag. & & 18.46 & 16.89 \\
%$L/L_\odot$ & & 1.22$\times$10$^7$ & 3.05$\times$10$^7$ \\
%$\mu_{e}$ & & 20.13$\pm$0.06 & 18.57$\pm$0.04 \\
%$r_e$ & & 49.1$\pm$0.5  & 50.4$\pm$0.5 \\
%$n$ & & 1.18$\pm$0.03 & 1.14$\pm$0.02 \\
%$b/a$ & & 0.964$\pm$0.005 & 0.930$\pm$0.007 \\
%PA [$^\circ$] & & -10$\pm$9 & 109$\pm$3 \\
%\hline 
%\tablecomments{The double S\'ersic $g$ profile is the best fitting profile.  (a) Errors on total magnitudes are dominated by the sky determination and are $< 0.05$ mag.}
%\end{deluxetable}

\subsection{Spectroscopy}

\subsubsection{Dynamical Mass, Mass-to-Light Ratio, and Resolved Kinematics}

Using the Keck/DEIMOS spectrum, the integrated velocity dispersion of M60-UCD1 was measured by cross-correlating the region around the Ca triplet with a library of templates of the same resolution and wavelength coverage, as described by Strader \etal~(2011).  This value is $\sigma_p=68\pm5$ \kms. The radial velocity of M60-UCD1 is $1290\pm5$ \kms; the systemic velocity of M60 is 1117 \kms (Gonz{\'a}lez 1993).

We estimate a dynamical mass for M60-UCD1 using the virial theorem: $M_{\rm vir}=\beta\sigma^{2}_{e} r_{e}/G$. $\beta$ is a parameter that depends on the structure of the galaxy and is smaller for more concentrated systems; $\sigma_e$ is the integrated velocity dispersion within $r_e$. Following the results of Cappellari \etal~(2006) for a range of S\'ersic profiles, we assume $\beta=7$, intermediate between the applicable values for the $n=3.3$ and $n=1.2$ profiles (corresponding to the inner and outer components respectively). We further estimate that $\sigma_e=71\pm5$ \kms, slightly higher than the measured value of  $\sigma_p$, by integrating over our DEIMOS extraction window (1.2\arcsec\ $\times$ 1.0\arcsec) and accounting for seeing.

The dynamical mass determined in this manner is $M_{\rm vir}=(2.0\pm0.3)\times10^8 M_{\odot}$. The systematic uncertainties are significant: we have assumed isotropy, sphericity, and 
mass-follows-light.

Dividing this dynamical mass by the total luminosity of M60-UCD1 yields a mass-to-light ratio of $M/L_V=4.9\pm0.7$. The flexible stellar population synthesis models of Conroy \etal~(2009), using Padova isochrones and a Kroupa initial mass function (IMF), predict $M/L_V=$ (3.5, 4.2, 4.7) for solar metallicity and ages of (8,10,12) Gyr, respectively\footnote{Using a Salpeter IMF gives $M/L_V=$ (5.9, 7.0, 7.9) for these ages, an increase of nearly 70\% in stellar mass over a Kroupa IMF.}. If M60-UCD1 has a younger age, the dynamical $M/L_V$ could imply an elevation in low-mass stars or stellar remnants over Kroupa IMF model predictions. For older ages there is an excellent match between the observed $M/L_V$ and the model predictions. As discussed in \S 3.3, a modest increase in the central velocity dispersion (and hence $M/L$) could also be caused by the presence of a supermassive black hole with a mass $\sim10$\% of that of the UCD (Mieske \etal~2013). Dark matter is not expected to contribute to the $M/L$ (\S 4.2).

M60-UCD1 is marginally resolved in our DEIMOS observations, and so some spatially-resolved kinematic information is available. The 5\arcsec\ slitlet was aligned close to the major axis of the object. Using the sky-subtracted two-dimensional spectrum, we determined the radial velocity and velocity dispersion on a pixel-by-pixel basis (one pixel is $\sim0.12$\arcsec). There is clear rotation present, with an amplitude of $\sim30$ km s$^{-1}$ to a projected radius of $\sim1.1$\arcsec. A decline of comparable amplitude in the velocity dispersion is also observed. Since the radial profiles are strongly affected by seeing, we do not attempt dynamical modeling. However, these data provide motivation to obtain improved kinematic maps in the future using integral-field spectroscopy.

\subsubsection{Abundances}

We constrain the stellar populations of M60-UCD1 using our MMT/Hectospec spectrum (with its wide wavelength range) and the models of Conroy \& van Dokkum (2012a; with additions from Conroy \& van Dokkum 2012b). These are stellar population synthesis models with variable abundance ratios for 11 elements. A Markov Chain Monte Carlo method is used to simultaneously fit the entire available optical spectrum.

The derived values are listed in Table 2. The uncertainties quoted are solely statistical, and do not include the substantial systematic uncertainties necessarily present in any integrated-light study of stellar populations. The rms residuals in the fit were $< 1$\% over most of the spectrum. The formal age is $14.5\pm0.5$ Gyr, indicating an old stellar population.

M60-UCD1 is of solar metallicity with a mild elevation in [$\alpha$/Fe] over solar. The abundances for C, O, and $\alpha$-elements appear very similar to the mean values for high-$\sigma$ local early-type galaxies determined in a similar manner (Conroy \etal~2013). However, the abundance of N is unusual: it is strongly enhanced, with [N/Fe$]\sim+0.6$, comparable to the average value observed in globular clusters (e.g., Briley \etal~2004). The high abundance of N in globular clusters is generally attributed to self-enrichment by the winds of asymptotic giant branch stars (Gratton \etal~2012). Our results suggest  light-element self-enrichment may also be present in UCDs, presumably related to their high stellar densities.

We note [Na/Fe] varies strongly with $\sigma$ in early-type galaxies, increasing from $\sim0$ to $\sim+0.4$ for $\sigma\sim140$ to $300$. The M60-UCD1 abundance ([Na/Fe$]\sim+0.4$) 
is therefore difficult to interpret: it could be expected, or could represent a large enhancement over baseline.

As a check on these values, we performed a standard Lick index analysis using {\tt EZ\_Ages} (Graves \& Schiavon 2008). These values are also listed in Table 2, and the same caveats apply. This analysis gave a formal age of $\sim9$--11 Gyr and similar abundance values for most of the elements in common with the full-spectrum analysis (C, Ca, Mg, Fe). For N the {\tt EZ\_Ages} analysis does not yield a reliable value, as the CN index strength is off the grids. The Lick CN$_2$ index for M60-UCD1 (0.24 mag) is comparable to that in most metal-rich M31 globular clusters, which are also thought to be self-enriched (Schiavon \etal~2012).

The very high abundance of N appears to be a robust conclusion of the analysis.

\begin{deluxetable}{lcc}
%\tablewidth{0pt}
\tablecaption{M60-UCD1 Abundances \label{tab:dat2}}
\tablehead{Element & Full Spec. & Lick \\
& (dex) & (dex)}

\startdata
[Fe/H] & $-0.02\pm0.02$  & $+0.06\pm0.03$ \\

[O/Fe] & $+0.19\pm0.07$ & \nodata \\

[C/Fe] &  $+0.10\pm0.03$ & $+0.02\pm0.04$ \\

[N/Fe] & $+0.61\pm0.04$ & \nodata \\

[Na/Fe] & $+0.42\pm0.03$\tablenotemark{a} & \nodata  \\

[Mg/Fe] & $+0.22\pm0.02$ & $+0.26\pm0.03$ \\

[Si/Fe] & $+0.12\pm0.05$ & \nodata \\

[Ca/Fe] & $+0.03\pm0.02$ & $-0.01\pm0.02$ \\

[Ti/Fe] & $+0.16\pm0.03$ &  \nodata \\
\enddata
\tablenotetext{a}{This abundance is largely determined by the resonance doublet at 589 nm, but the 819nm \ion{Na}{1} line gives a consistent result.}
\end{deluxetable}

\subsection{X-ray Data: Central Black Hole or X-ray Binary?}

An X-ray source at the position of M60-UCD1 is present in the \emph{Chandra}/ACIS catalog of Luo \etal~(2013). The central astrometric matching between the $Chandra$ and $HST$ data is excellent due to the large X-ray binary and globular cluster populations of M60, with an rms scatter of 0.17\arcsec.

This X-ray source, termed XID 144 by Luo \etal~(2013), has a position consistent with the optical center of M60-UCD1. There is evidence that it is variable, with its X-ray luminosity (from 0.3 to 8 keV) ranging from $\sim6\times10^{37}$ to $\sim1.3\times10^{38}$ erg/s over timescales as short as a few months. The X-ray spectrum is well-fit by a absorbed power-law with a photon index of 1.8.

There are two reasonable possibilities for this central X-ray source: it could either be an active galactic nucleus associated with a massive black hole or a low-mass X-ray binary.

The case for a central black hole is straightforward. If the black hole occupation fraction of dwarf galaxies is high, and if  UCDs are the products of tidal stripping of dwarf galaxies, then a significant fraction of UCDs should have ``overmassive" black holes that could be detected through dynamical or accretion signatures. If UCDs have been stripped of 99\% or more of their original mass (we estimate in \S 4.2 that the progenitor of M60-UCD1 was $\sim50$--200 times more massive), then they could host supermassive black holes that are $\ga10$\% of their present-day masses (Mieske \etal~2013). Frank \etal~(2011) constrain a putative black hole to be $< 5\%$ of the total mass of one Fornax cluster UCD through integral-field spectroscopy.

The observed X-ray luminosity would be consistent with a $\sim10^{7} M_{\odot}$ black hole accreting at $10^{-4}$ of the Eddington rate with a radiative efficiency of $10^{-3}$. This Eddington ratio of $10^{-7}$ would be typical of nuclei with old stellar populations (Ho 2009).

We can also estimate the odds that M60-UCD1 contains a bright X-ray binary. Sivakoff \etal~(2007) derive formulae to estimate the probability that a globular cluster contains a low-mass X-ray binary with $L_X > 3.2\times10^{38}$ erg/s. The odds are higher for metal-rich clusters and those with high encounter rates. Applying their results, but extrapolating to the fainter luminosity observed, suggests a $\sim25$\% chance of having observed a low-mass X-ray binary in M60-UCD1. However, these results are of uncertain relevance for an object with a different structure and formation history than a globular cluster (since what is pertinent is the integrated---not instantaneous---collision rate). Dabringhausen \etal~(2012) suggest UCDs have a lower occurrence of low-mass X-ray binaries than expected on the basis of the Sivakoff \etal~(2007) results.

Future observations can help clarify the nature of the X-ray source. For example, if M60-UCD1 hosts a $\ga10^6 M_{\odot}$ black hole that lies on the radio--X-ray fundamental plane for black holes (Plotkin \etal~2012), it should be detectable with the Very Large Array.

\section{Discussion}

\subsection{The Densest Galaxy?}

Figure 3 shows a  plot of log $\Sigma$ vs.~log $L_V$  for dispersion-supported stellar systems.  $\Sigma$ is the mean surface luminosity density within $r_e$.
Globular clusters are plotted with different symbols than galaxies. It is clear that M60-UCD1 is an unusual object: it is much denser than any other object classified as a galaxy. It is more massive than any UCD or star cluster of comparable size, but is much more compact than other galaxies of similar luminosity. 

\begin{figure}
\begin{center}
\epsscale{1.2}
%\plotone{e2.eps}
\plotone{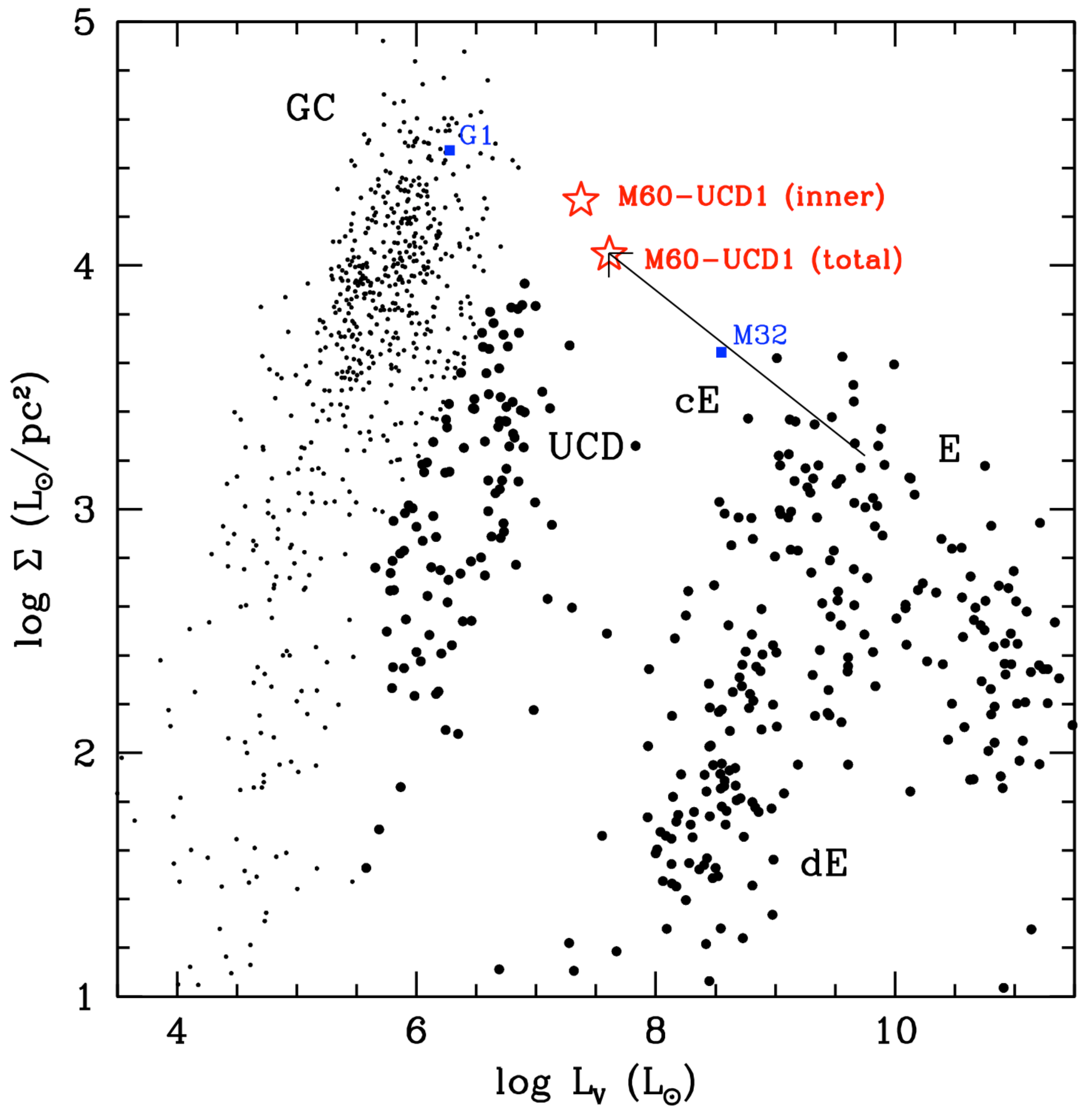}
\caption{log $\Sigma$ (mean surface luminosity density within $r_e$) vs.~log $L_V$ for dispersion-supported stellar systems (GC=globular cluster; cE=compact elliptical; E=early-type galaxy; dE=dwarf elliptical). The inner component and overall parameters for M60-UCD1 (red stars) are marked, as are the comparison objects M32 and the luminous M31 cluster G1 (blue squares). Globular clusters (the union of objects with $r_e<10$ pc and non-dwarf galaxies with $M_V>-9$; Brodie \etal~2011) are small points; galaxies are large points. M60-UCD1 has a higher $\Sigma$ than any other galaxy. The black arrow represents the proposed evolution of the progenitor of M60-UCD1 as it was tidally stripped. The principal data source for this figure is the spectroscopically-confirmed compilation of Brodie \etal~(2011)\footnote{See http://sages.ucolick.org/downloads/sizetable.txt}, with updates from Forbes \etal~(2013).}

\label{ucd}
\end{center}
\end{figure}

M60-UCD1 is not the densest stellar system known. That honor goes to any of a number of nuclear star clusters, which can reach mean surface densities of $>10^{5}M_{\odot}$ pc$^{-2}$ within $r_e$ (Walcher \etal~2005; these are not plotted in Figure 3). Many massive globular clusters are also extremely dense. However, M60-UCD1 is arguably the densest \emph{galaxy} known in the local universe. Using the $M/L_V$ from \S3.2.1, its mean effective surface density is $\Sigma=5.4\times10^{4}M_{\odot}$ pc$^{-2}$, a factor of 2.5--3 higher than for M32. The inner component of M60-UCD1, with $r_e\sim14$ pc, has a mean $\Sigma\sim9\times10^{4}M_{\odot}$ pc$^{-2}$, comparable to that of many nuclear star clusters. The central volume density of M60-UCD1 is not well-constrained by the present data.

The object most similar to M60-UCD1 is HUCD1, a Hydra Cluster UCD, which has $r_e=25$ pc and $M_V=-13.4$ (Misgeld \etal~2011), though M59cO (Chilingarian \& Mamon 2008) and several Coma Cluster UCDs (Chiboucas \etal~2011) are also similar, if less extreme. It seems likely that ongoing surveys for UCDs will turn up additional objects with  properties comparable to M60-UCD1.

\subsection{The Origin of M60-UCD1}

The extreme mass, multiple structural components, high metallicity, and possible presence of a central black hole make it unlikely that M60-UCD1 is a star cluster or merged cluster complex. It is most plausible that the object is the tidally-stripped remnant of a more massive progenitor galaxy. 

Pfeffer \& Baumgardt (2013) present new simulations of the formation of UCDs through tidal stripping of nucleated galaxies in a massive, Virgo-like cluster. They show that it is possible to reproduce the sizes and luminosities of typical UCDs. These simulations were not intended to match the most massive UCDs, and hence are not directly comparable to the properties of M60-UCD1. However, a general conclusion from this study is relevant: the nucleus is largely unaffected by the stripping process, so the inner core of the resultant UCD should have similar properties to the original nucleus. 

The inner component of M60-UCD1 has $M_g=-13.1$, $g-z=1.54$, and $r_e=14$ pc. The size, luminosity, and red color are similar to nuclei in the Virgo galaxies NGC 4379 and NGC 4387 (C{\^o}t{\'e} \etal~2006) and the Fornax Cluster galaxies NGC 1389 and IC 2006 (Turner \etal~2012). These galaxies have $M_B\sim-18$ to $-19$ and stellar masses $\sim1$--$3\times10^{10} M_{\odot}$. We conclude that M60-UCD1 could have originated in the tidal stripping of a Virgo galaxy in this luminosity range. Since the present luminosity of M60-UCD1 is $M_B\sim-13.2$, we infer that it is a factor of $\sim50$--200 less massive because of the stripping. The projected distance of M60-UCD1 from the center of M60 ($\sim6.6$ kpc) is consistent with the small pericenter needed for efficient stripping (Pfeffer \& Baumgardt 2013). Galaxies in this mass range host significant globular cluster populations ($\sim40$--100; Brodie \& Strader 2006) that would also be stripped during UCD formation and might still be detectable in phase space (e.g., Romanowsky \etal~2012).

We caution that UCDs may originate in a biased subset of host galaxies that have been largely destroyed, so it is possible that there is no correspondence between UCD progenitors  and a subpopulation of surviving galaxies.

The Pfeffer \& Baumgardt (2013) simulations show that extended debris suggestive of tidal stripping becomes challenging to observe after relatively short ($\sim1$ Gyr) timescales, so the lack of evidence for such debris in Figure 2 does not disfavor this scenario. The dynamical friction timescale (Binney \& Tremaine 1987) for M60-UCD1 is $\sim5$ Gyr, so it is plausible that its progenitor was stripped long ago and the remnant has ``stalled" at its current radius. In this case no observable tidal tails would be expected.

The stellar density of M60-UCD1 is much higher than expected for dark matter with a standard NFW profile (a factor of $\sim15$ for even for extreme, cluster-scale halo masses; Tollerud \etal~2011). If the galaxy has undergone as much tidal stripping as inferred, this is likely to have strongly modified the dark matter profile; nonetheless, M60-UCD1 is probably the worst UCD in which to search for dark matter. As discussed by Willman \& Strader (2012; see also Hilker \etal~2007), the UCDs most likely to show evidence for dark matter are the least massive and most extended UCDs.

Future observations will help constrain the detailed properties of M60-UCD1, including its two-dimensional kinematics and whether it hosts a central supermassive black hole. 

\acknowledgments

We thank L.~Chomiuk, S.~Mieske, and R.~Schiavon for useful discussions and an anonymous referee for a helpful report. Based on observations made with the NASA/ESA \emph{Hubble Space Telescope} and the Hubble Legacy Archive. Data obtained at Keck (Caltech/UC/NASA) and MMT (Arizona/Smithsonian). Products produced by the OIR Telescope Data Center (SAO). Support by ARC grant DP130100388 and NSF grants AST-1109878/AST-0909237.


\begin{thebibliography}{}

\bibitem[Abazajian et al.(2009)]{2009ApJS..182..543A} Abazajian, K.~N., Adelman-McCarthy, J.~K., Ag{\"u}eros, M.~A., et al.\ 2009, \apjs, 182, 543 
\bibitem[Binney \& Tremaine(1987)]{1987gady.book.....B} Binney, J., \& Tremaine, S.\ 1987, Princeton, NJ, Princeton University Press
\bibitem[Briley et al.(2004)]{2004AJ....127.1588B} Briley, M.~M., Harbeck, D., Smith, G.~H., \& Grebel, E.~K.\ 2004, \aj, 127, 1588 
\bibitem[Blakeslee et al.(2009)]{2009ApJ...694..556B} Blakeslee, J.~P., Jord{\'a}n, A., Mei, S., et al.\ 2009, \apj, 694, 556 
\bibitem[Brodie et al.(2011)]{2011AJ....142..199B} Brodie, J.~P., Romanowsky, A.~J., Strader, J., \& Forbes, D.~A.\ 2011, \aj, 142, 199 
\bibitem[Brodie \& Strader(2006)]{2006ARA&A..44..193B} Brodie, J.~P., \& Strader, J.\ 2006, \araa, 44, 193 
\bibitem[Br{\"u}ns et al.(2011)]{2011A&A...529A.138B} Br{\"u}ns, R.~C., Kroupa, P., Fellhauer, M., Metz, M., \& Assmann, P.\ 2011, \aap, 529, A138 
\bibitem[Cappellari et al.(2006)]{2006MNRAS.366.1126C} Cappellari, M., Bacon, R., Bureau, M., et al.\ 2006, \mnras, 366, 1126 
\bibitem[Chiboucas et al.(2011)]{2011ApJ...737...86C} Chiboucas, K., Tully, R.~B., Marzke, R.~O., et al.\ 2011, \apj, 737, 86 
\bibitem[Chilingarian \& Mamon(2008)]{2008MNRAS.385L..83C} Chilingarian, I.~V., \& Mamon, G.~A.\ 2008, \mnras, 385, L83 
\bibitem[Chilingarian et al.(2011)]{2011MNRAS.412.1627C} Chilingarian, I.~V., Mieske, S., Hilker, M., \& Infante, L.\ 2011, \mnras, 412, 1627 
\bibitem[Conroy et al.(2009)]{2009ApJ...699..486C} Conroy, C., Gunn, J.~E., \& White, M.\ 2009, \apj, 699, 486 
\bibitem[Conroy \& van Dokkum(2012a)]{2012ApJ...747...69C} Conroy, C., \& van Dokkum, P.\ 2012, \apj, 747, 69 
\bibitem[Conroy \& van Dokkum(2012b)]{2012ApJ...760...71C} Conroy, C., \& van Dokkum, P.~G.\ 2012, \apj, 760, 71 
\bibitem[Conroy et al.(2013)]{2013arXiv1303.6629C} Conroy, C., Graves, G., \& van Dokkum, P.\ 2013, ApJ, submitted (arXiv:1303.6629)
\bibitem[Cooper et al.(2012)]{2012ascl.soft03003C} Cooper, M.~C., Newman, J.~A., Davis, M., Finkbeiner, D.~P., \& Gerke, B.~F.\ 2012, Astrophysics Source Code Library, 3003 
\bibitem[C{\^o}t{\'e} et al.(2004)]{2004ApJS..153..223C} C{\^o}t{\'e}, P., Blakeslee, J.~P., Ferrarese, L., et al.\ 2004, \apjs, 153, 223 
\bibitem[C{\^o}t{\'e} et al.(2006)]{2006ApJS..165...57C} C{\^o}t{\'e}, P., Piatek, S., Ferrarese, L., et al.\ 2006, \apjs, 165, 57 
\bibitem[Dabringhausen et al.(2012)]{2012ApJ...747...72D} Dabringhausen, J., Kroupa, P., Pflamm-Altenburg, J., \& Mieske, S.\ 2012, \apj, 747, 72 
\bibitem[Drinkwater et al.(2000)]{2000PASA...17..227D} Drinkwater, M.~J., Jones, J.~B., Gregg, M.~D., \& Phillipps, S.\ 2000, PASA, 17, 227 
\bibitem[Drinkwater et al.(2003)]{2003Natur.423..519D} Drinkwater, M.~J., Gregg, M.~D., Hilker, M., et al.\ 2003, \nat, 423, 519 
\bibitem[Faber et al.(2003)]{2003SPIE.4841.1657F} Faber, S.~M., Phillips, A.~C., Kibrick, R.~I., et al.\ 2003, \procspie, 4841, 1657 
\bibitem[Fabricant et al.(2005)]{2005PASP..117.1411F} Fabricant, D., Fata, R., Roll, J., et al.\ 2005, \pasp, 117, 1411 
\bibitem[Forbes et al.(2013)]{2013arXiv1306.5245F} Forbes, D., Pota, V., Usher, C., et al.\ 2013, MNRAS, in press (arXiv:1306.5245)
\bibitem[Frank et al.(2011)]{2011MNRAS.414L..70F} Frank, M.~J., Hilker, M., Mieske, S., et al.\ 2011, \mnras, 414, L70 
\bibitem[Gonz{\'a}lez(1993)]{1993PhDT.......172G} Gonz{\'a}lez, J.~J.\ 1993, Ph.D.~Thesis, University of California, Santa Cruz
\bibitem[Gratton et al.(2012)]{2012A&ARv..20...50G} Gratton, R.~G., Carretta, E., \& Bragaglia, A.\ 2012, \aapr, 20, 50 
\bibitem[Graves \& Schiavon(2008)]{2008ApJS..177..446G} Graves, G.~J., \& Schiavon, R.~P.\ 2008, \apjs, 177, 446 
\bibitem[Hilker et al.(1999)]{1999A&AS..134...75H} Hilker, M., Infante, L., Vieira, G., Kissler-Patig, M., \& Richtler, T.\ 1999, \aaps, 134, 75 
\bibitem[Hilker et al.(2007)]{2007A&A...463..119H} Hilker, M., Baumgardt, H., Infante, L., et al.\ 2007, \aap, 463, 119 
\bibitem[Ho(2009)]{2009ApJ...699..626H} Ho, L.~C.\ 2009, \apj, 699, 626 
\bibitem[Luo et al.(2013)]{2013ApJS..204...14L} Luo, B., Fabbiano, G., Strader, J., et al.\ 2013, \apjs, 204, 14 
\bibitem[Mieske et al.(2013)]{2013mieske} Mieske, S., Frank, M., Baumgardt, H., Luetzgendorf, N., Neumayer, N., Hilker, M. 2013, \aap, in press (arXiv: 1308.1398)
\bibitem[Mieske et al.(2012)]{2012A&A...537A...3M} Mieske, S., Hilker, M., \& Misgeld, I.\ 2012, \aap, 537, A3 
\bibitem[Mink et al.(2007)]{2007ASPC..376..249M} Mink, D.~J., Wyatt, W.~F., Caldwell, N., et al.\ 2007, Astronomical Data Analysis Software and Systems XVI, 376, 249 
\bibitem[Misgeld et al.(2011)]{2011A&A...531A...4M} Misgeld, I., Mieske, S., Hilker, M., et al.\ 2011, \aap, 531, A4
\bibitem[Norris \& Kannappan(2011)]{2011MNRAS.414..739N} Norris, M.~A., \& Kannappan, S.~J.\ 2011, \mnras, 414, 739 
\bibitem[Pfeffer \& Baumgardt(2013)]{2013MNRAS.tmp.1504P} Pfeffer, J., \& Baumgardt, H.\ 2013, \mnras, 1504 
\bibitem[Plotkin et al.(2012)]{2012MNRAS.419..267P} Plotkin, R.~M., Markoff, S., Kelly, B.~C., K{\"o}rding, E., \& Anderson, S.~F.\ 2012, \mnras, 419, 267 
\bibitem[Romanowsky et al.(2012)]{2012ApJ...748...29R} Romanowsky, A.~J., Strader, J., Brodie, J.~P., et al.\ 2012, \apj, 748, 29 
\bibitem[Schiavon et al.(2012)]{2012AJ....143...14S} Schiavon, R.~P., Caldwell, N., Morrison, H., et al.\ 2012, \aj, 143, 14 
\bibitem[Seth et al.(2006)]{2006AJ....132.2539S} Seth, A.~C., Dalcanton, J.~J., Hodge, P.~W., \& Debattista, V.~P.\ 2006, \aj, 132, 2539 
\bibitem[Simard et al.(2011)]{2011ApJS..196...11S} Simard, L., Mendel, J.~T., Patton, D.~R., Ellison, S.~L., \& McConnachie, A.~W.\ 2011, \apjs, 196, 11 
\bibitem[Sivakoff et al.(2007)]{2007ApJ...660.1246S} Sivakoff, G.~R., Jord{\'a}n, A., Sarazin, C.~L., et al.\ 2007, \apj, 660, 1246 
\bibitem[Strader et al.(2011)]{2011AJ....142....8S} Strader, J., Caldwell, N., \& Seth, A.~C.\ 2011, \aj, 142, 8 
\bibitem[Strader et al.(2012)]{2012ApJ...760...87S} Strader, J., Fabbiano, G., Luo, B., et al.\ 2012, \apj, 760, 87 
\bibitem[Tollerud et al.(2011)]{2011ApJ...726..108T} Tollerud, E.~J., Bullock, J.~S., Graves, G.~J., \& Wolf, J.\ 2011, \apj, 726, 108 
\bibitem[Turner et al.(2012)]{2012ApJS..203....5T} Turner, M.~L., C{\^o}t{\'e}, P., Ferrarese, L., et al.\ 2012, \apjs, 203, 5 
\bibitem[Walcher et al.(2005)]{2005ApJ...618..237W} Walcher, C.~J., van der Marel, R.~P., McLaughlin, D., et al.\ 2005, \apj, 618, 237 
\bibitem[Willman \& Strader(2012)]{2012AJ....144...76W} Willman, B., \& Strader, J.\ 2012, \aj, 144, 76 


\end{thebibliography}
\end{document}